\documentclass[10pt,journal]{IEEEtran}

\ifCLASSINFOpdf
   \usepackage[pdftex]{graphicx}
   \usepackage{xcolor}
\else
\fi

\usepackage{multirow}
\usepackage{amsmath,amsfonts}
\usepackage[dvipsnames]{xcolor}
\usepackage{biblatex}
\usepackage{listingsutf8}
\usepackage{xurl}
\usepackage{fontawesome5}
\usepackage{siunitx}
\usepackage{paralist} 
\usepackage[draft]{changes}

\usepackage{lipsum}

\def\BibTeX{{\rm B\kern-.05em{\sc i\kern-.025em b}\kern-.08em
    T\kern-.1667em\lower.7ex\hbox{E}\kern-.125emX}}
    
\newcommand{\iic}{I\textsuperscript{2}C}

\begin{document}

\title{A UEFI System with SPDM to Protect Against Unauthorized Device Connections}

\author{Ágatha de Freitas,
        Marcos A. Simplicio Jr,
        Bruno C. Albertini,
        Renan C. A. Alves%
\thanks{The authors are with Universidade de São Paulo, Brazil.}
}



\maketitle

\begin{abstract}
Attackers willing to compromise computing systems can use malicious peripherals as an attack vector, threatening users that cannot verify the hardware's authenticity.
To address this problem, our work uses the Security Protocol and Data Model to propose a UEFI system capable of authenticating PCIe and USB devices trying to connect with it.
We also develop an open source proof-of-concept using emulation to evaluate and illustrate our proposal, which is capable of restricting the devices' connections to only those allowed, thus protecting the system against malicious peripherals.
Then, using kernel virtualization features to evaluate the emulation, we collect the number of instructions and CPU cycles during boot.
Our experiments reveal that, during firmware execution, the number of instructions and the number of CPU cycles increased respectively 13\% and 8\% on average.
This processing overhead is acceptable in view of enhanced security.
Institutions requiring high security levels can leverage our proof-of-concept to tailor their own system based on their own requirements.
\end{abstract}


\IEEEpeerreviewmaketitle

\section{Introduction} \label{sec:introduction}

\IEEEPARstart{I}{n} modern computer systems, the traditional boot method depends on a pre-installed Basic Input/Output System (BIOS) which is being replaced by Extensible Firmware Interface (EFI).
This replacement is happening as the BIOS is not standardized among companies, leading to multiple firmware implementations requiring patches to support new emerging features.
Additionally, the lack of standardization forces developers who need the firmware interface to thoroughly understand the component's operation.
Observing these problems, Intel and Microsoft proposed the EFI in 2004~\cite{introducing-efi:2004}, offering better boot speed, extensibility, and security when comparing to BIOS in order to match computer evolution~\cite{nsa-uefi:2018}.
It also motivated the creation of the Unified EFI (UEFI), which aims to standardize the firmware interface and decrease the developers' burden as they handle new technologies.

When correctly configured, UEFI firmware establishes a trust chain for the operating system (OS), allowing only the execution of signed binaries when their hashes are matched with an allowlist~\cite{nsa-uefi:2018,nsa-boot-modes:2019}.
On the other hand, there are menaces for incorrectly configured systems that are able to persist even after OS reinstallation or hard drive replacement (e.g, MosaicRegressor, a espionage toolkit, and LoJax, a toolkit to apply update patches on flash memories holding the firmware).
Such characteristics of these UEFI threats furnish attackers the possibility of building complex attack chains to compromise the firmware and, consequently, the entire system.
These threats gain relevance as organizations requiring high security standards, such as government agencies, financial institutions, and companies with valuable intellectual properties, engage on commercial agreements with stakeholders scattered around the world to acquire and develop hardware components.
Therefore, UEFI persistent attacks may come from ill intentioned members of the supply chain, which can silently tamper with the shipped devices~\cite{real-world-hardware-trojan:2023}, or an external device trying to connect with the target, e.g., Stuxnet compromised a nuclear plant with a malicious USB flash drive\footnote{\url{https://www.rusi.org/news-and-comment/in-the-news/how-simple-usb-stick-sabotaged-irans-nuclear-plan-world-first-showdown}}.

Addressing hardware and firmware security, Distributed Management Task Force (DMTF) specifies a broad and promising solution to protect a computer system from such threats named as Security Protocol and Data Model (SPDM)~\cite{spdm-1.3.2:2024}.
It defines messages, data objects, and sequences providing means to authenticate devices and measure their firmwares, however companies are still in the process of designing and producing SPDM-enabled hardware (e.g., AMD, Nvidia, HPE, and Intel).
Due to the lack of hardware support, the protocol has been being implemented on the OS layer~\cite{NVIDIA_drivers,ZeroTrust:2023,enhancing-circuit-auth:2024,securing-hard-drives:2022,testing-the-limits:2025}, consequently firmware initialization is not protected.
%

\section{Proposal} \label{sec:proposal}

This work aims to propose a computer system design that uses UEFI firmware with secure boot enabled and is capable of authenticating and measuring PCIe and USB devices with SPDM.
Additionally, this work intends to propose an architecture to authenticate and measure all USB devices with distinct transfer types, for example, control, bulk, isochronous, and interrupt transfers.
Thus, the goal system is a reference of how to use the SPDM standard to protect itself from malicious devices willing to stablish connection during boot.
We expect this work to promote the use of SPDM in real scenarios -- specially in environments demanding high security levels, such as financial institutions and government agencies -- as the design can be tailored to different scenarios (e.g., authenticate devices using different buses, adapt our PoC to computer systems with different architectures, etc.).
%

\subsection{System Requirements} \label{sec:system-requirements}

Since this work aims to contribute to the community, it is necessary to choose open source tools with freely available documentation.
Also, the communities of these tools must be active, proving the viability of implementing our design in real scenarios due to these softwares reach.
Therefore, regarding emulation, we choose QEMU~\cite{qemu:2005} as it attends aforementioned requirements.
Although this emulator cannot use the security standard immediately, it is possible to enable this feature by inserting the libspdm~\cite{libspdm} inside QEMU's code tree, then it is possible to develop a component supporting SPDM.
The firmware, on the other hand, must be compliant to the UEFI specification and also be compatible to the selected emulator.
Facing this requirement, we select the Open Virtual Machine Firmware (OVMF)~\cite{tianocore-ovmf:2017}, for it has been developed based on the EFI Development Kit II (EDK II), which the TianoCore community maintains as an Intel initiative, for QEMU.

\begin{figure}[!tb]
    \centering
    \includegraphics[width=\linewidth]{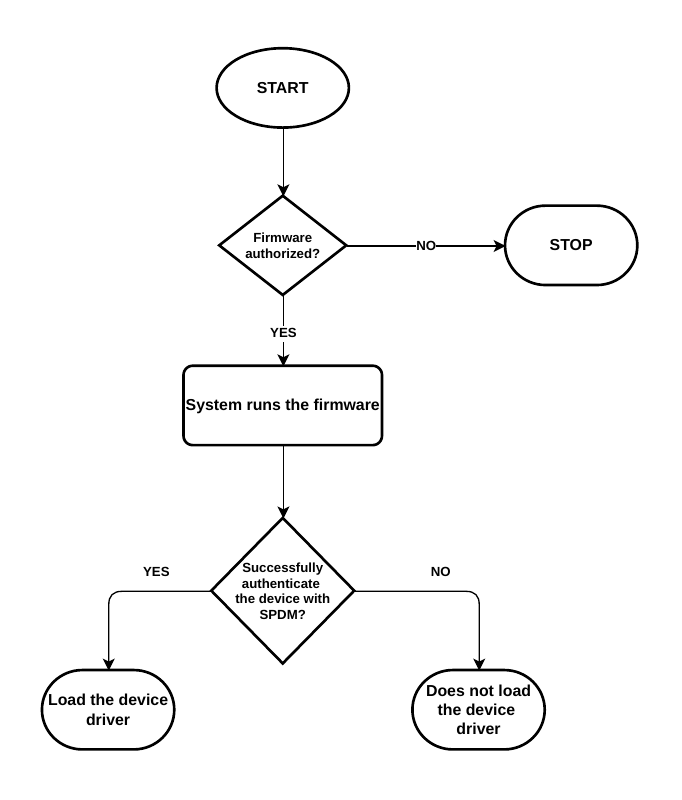}
    \caption{System behavior of secure boot with SPDM.}
    \label{fig:flow-boot-spdm}
\end{figure}

Concerning the system requirements, at the moment the platform is turned on, it must firstly verify the firmware inside the flash memory, and, if it is not tampered, it may be executed; if by any means the firmware suffered modifications, the emulation must stop.
Afterwards, the OVMF must authenticate and measure the PCIe and USB devices connected, further, they must not trust the firmware identity, requiring them to perform mutual authentication.
Albeit this method leads to multiple authentications targeting OVMF, this behavior is desired since, in our design, no device must assume they can trust the firmware.
Besides the binary verification in the flash memory, a failure in authentication or measurement of a device must not terminate the emulation, but its usage must not be allowed during the entire firmware execution.
Fig.~\ref{fig:flow-boot-spdm} shows the system target behavior.

\section{Attacker Model} \label{sec:attacker-model}

We define the attacker our system aims to inhibit using OWASP ISTG~\cite{owasp-istg:2024} as it covers general hardware and firmware security topics.
Therefore, the attacker must have physical access (PA) level 3 or 4, i.e., they must have physical access to the machine, whereas the former level does not have access to the internal components and the latter one has.
Concerning authorization access level (AA), the attacker can have any level depending on the target system (e.g., unauthorized access, low-privileged access, high-privileged access, and manufacturer-level access, in ascending level order).

\subsection{Physical Access Level 3} \label{subsec:pa3}

An employee with physical access to a computer system furnished by the employer is able to connect a device that may be malicious even if they do not know about it.
Depending on the security configuration of the system, the contents of this connected device is able to infect the host with malware (e.g., bootkits, rootkits, ransomware, etc.).
Another potential threat is an attacker with access to the supply chain who is able to stealthily replace a device being shipped by a malicious one, which would be considered authentic unless the target system has a way to verify it.
Finally, there is also the possibility of an attacker pretending to be an authentic user of a business service with physical access to the company hardwares, who, in some way, is able to connect a malicious device to it.
All cases mentioned are PA-3, while the required access levels for a successful attack are AA-2 or AA-3 for the first two cases and AA-1 for the last one.

\subsection{Physical Access Level 4} \label{subsec:pa4}

A malicious employee with access to the assembly line, i.e., AA-4, and sufficient knowledge about the system can replace an authentic component by a malicious one.
However, even if the attack is PA-4 and AA-4, we assume the attacker cannot access the system private key, otherwise any authentication attempt is irrelevant to the system security.
For example, an employee in a laptop production line could replace the soldered drive by a tampered one as well as solder an additional spy hardware.
%

\section{Related Work} \label{sec:related-work}

There are few published works using the SPDM standard to authenticate and measure devices~\cite{ZeroTrust:2023,enhancing-circuit-auth:2024,securing-hard-drives:2022,testing-the-limits:2025}.
None of them uses SPDM in UEFI firmware on PCIe and USB buses to propose a secure platform initialization (PI).

While \cite{ZeroTrust:2023} and \cite{enhancing-circuit-auth:2024} use SPDM, the messages are not processed by the hardware itself, for the software runs the standard and routes the packets to the \iic bus via socket.
Even though our work implements the SPDM in emulated devices, they emulate a device capable of exchanging messages in accord to the standard without the need of another software to process the packets, i.e., the emulated hardware can answer the firmware driver directly.
Our proposal uses the SPDM to detect and disable the malicious components, similar to \cite{ZeroTrust:2023} and \cite{enhancing-circuit-auth:2024}, moreover, our system aims security since boot in UEFI machines with the help of a dedicated security module.

Other published works also have implementations in emulated environment~\cite{securing-hard-drives:2022,testing-the-limits:2025}, however the authors change the input and output functions of each device.
This technique inhibits scalability because it requires knowledge of each components' behavior.
In contrast, our proposal leverages the devices connections (e.g., PCIe and USB), developing generic functions able to use the buses to exchange SPDM messages, decreasing the required knowledge about the operation of each component.
Furthermore, \cite{securing-hard-drives:2022} and \cite{testing-the-limits:2025} modify the drivers in Linux kernel, thus offering protection after PI.

Besides published works, the TianoCore community has a repository with features that are not ready for inclusion in EDK II's main code tree, which contains EFI modules using SPDM to authenticate and measure PCIe devices~\cite{edk2-devicesecurity}.
The developers test these modules using the EDK II's own emulator by running them as applications in UEFI shell, i.e., component authentication only happens after the user interacts at least once with the system.
Nevertheless, in a UEFI firmware with secure boot enabled, the shell is not available for security reasons, turning this method unfeasible.
Therefore, this work leverages the available codes in \cite{edk2-devicesecurity} to develop the proof-of-concept (PoC) of our proposal, updating their code to make it compatible to current SPDM code inside EDK II.
Our proposed system also utilizes another implementation, not accepted in the EDK II's code tree as the maintainers considered it as a feature only useful for SPDM at the moment~\cite{edk2-pci-doe}.
We still use this pull request because \cite{edk2-devicesecurity} defines stub functions to simulate PCIe, which replicates the logic of read and write in its configuration space, while \cite{edk2-pci-doe} implements PCI DOE as an EDK II protocol using PCIe resources.
Our work differs from \cite{edk2-devicesecurity} on the following aspects:
\begin{enumerate}
    \item the driver with SPDM support is placed in the firmware binary at compilation time;
    \item the SPDM responder simulated by an EFI application is discarded;
    \item USB devices are authenticated and measured;
    \item we use an emulator, i.e., QEMU, capable of initializing multiple OSes, while the EDK II emulator can only initialize the EFI environment; and
    \item even though \cite{edk2-devicesecurity} can implement mutual authentication using SPDM, it does not do it, as crucial functions to enable it are not yet implemented.
\end{enumerate}

\cite{deviceveil:2019} describes a technique called DeviceVeil to individually authenticate USB devices.
This method isolates the peripheral in a host OS and authenticate it using the Physical Unclonable Function (PUF) module inside the device.
DeviceVeil is suited for OSes, i.e., it trusts the firmware which differs from our work, and the provided authentication mechanism is unilateral, hence it requires another method to provide mutual authentication, while SPDM itself provides this feature.

Table~\ref{tab:work-comparisons} summarizes related works comparing their characteristics with our proposal.

\begin{table}[!tb]
    \centering
    \caption{Comparison between our work and related ones.}
    \begin{tabular}{|c|c|c|c|}
        \hline
        & \multirow{3}{*}{\centering \textbf{UEFI}} & \multirow{3}{7em}{\centering \textbf{Mutual\\ Authentication}} & \multirow{3}{5.5em}{\centering \textbf{Emulated Computer System}} \\
        &&&\\
        &&&\\
        \hline
        \cite{ZeroTrust:2023} & \faTimes  & \faCheck & \faTimes \\
        \hline
        \cite{enhancing-circuit-auth:2024} & \faTimes  & \faCheck & \faTimes \\
        \hline
        \cite{securing-hard-drives:2022} & \faTimes  & \faCheck & \faCheck \\
        &&&\\
        \hline
        \cite{testing-the-limits:2025} & \faTimes & \faCheck & \faCheck \\
        \hline
        \cite{edk2-devicesecurity} & \faCheck & \faTimes & \faTimes \\
        \hline
        \cite{deviceveil:2019} & \faTimes & \faTimes & \faTimes \\
        \hline
        \textbf{Our work} & \faCheck & \faCheck & \faCheck \\
        \hline
    \end{tabular}
    \label{tab:work-comparisons}
\end{table}

\section{Background} \label{sec:background}

This section details the required concepts for the Section~\ref{sec:poc}.
It starts summarizing the SPDM standard in Section~\ref{subsec:spdm} and explains the resources of the security module we use in Section~\ref{subsec:tpm}.
%

\subsection{Security Protocol and Data Model} \label{subsec:spdm}

As stated in Section~\ref{sec:introduction}, the SPDM standard defines messages, objects, and sequences regardless of the communication protocols and devices~\cite{spdm-1.3.2:2024}.
It furnishes a way of verifying hardware identities through digital certificates and asymmetric keys, while their generation and protection are not specified.
In addition, it is possible to get firmware identity representations as well as configuration data, i.e., measurements, to check the device's integrity.
Besides those, SPDM can also establish a secure session to encrypt the communication contributing to confidentiality.

\begin{figure}[!tb]
  \begin{center}
    \includegraphics[width=\linewidth]{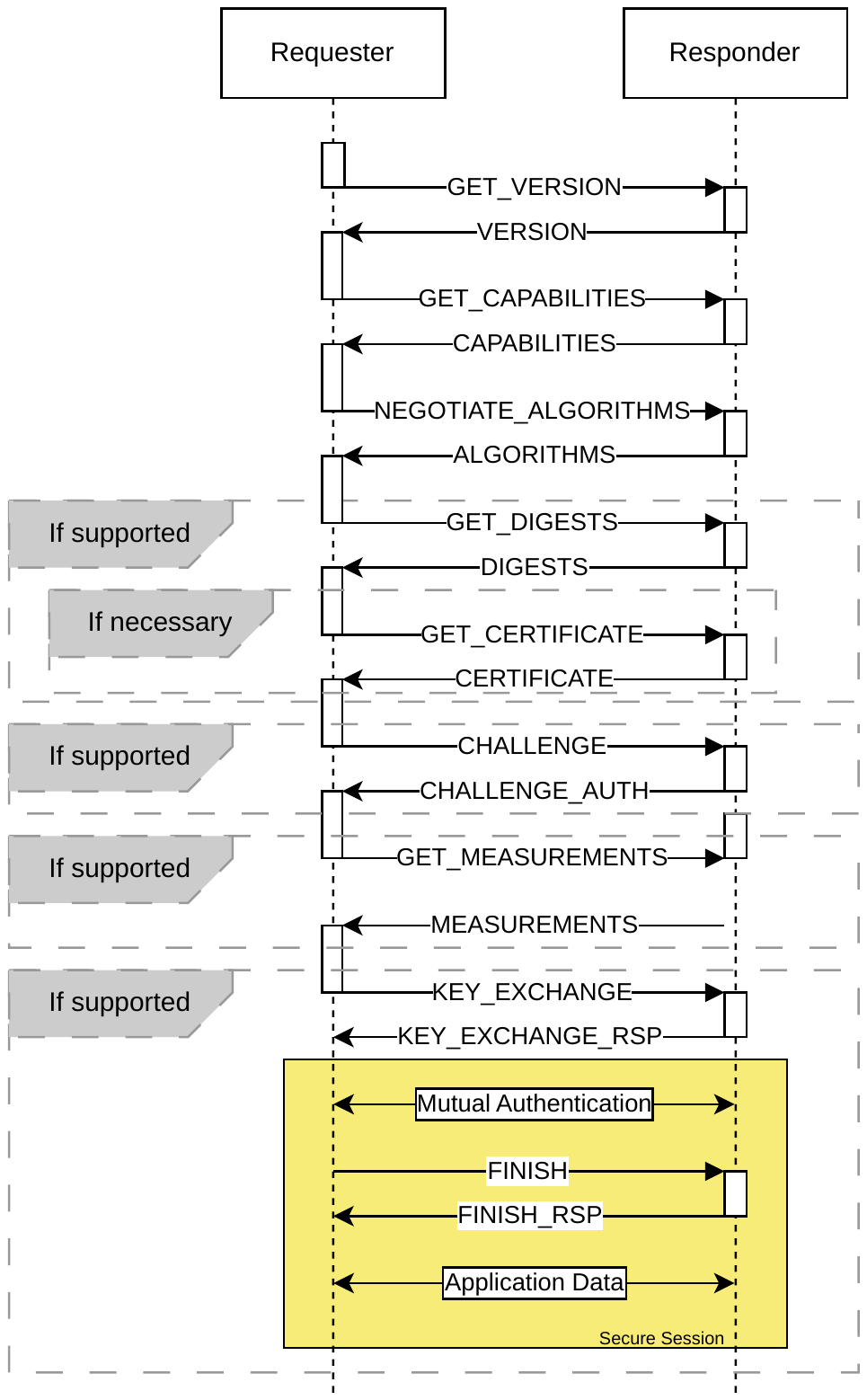}
  \end{center}
  \caption{SPDM communication flow. Adapted from \cite{spdm-1.3.2:2024}.}
  \label{fig:spdm-flow}
\end{figure}

As Fig.~\ref{fig:spdm-flow} shows in the complete message exchange flow, in an SPDM communication, a device can assume the following roles: requester, responder or both.
The standard allows multiple connections on a single endpoint, and they are managed separately, however the same device cannot assume both roles on the same connection.
Regarding their roles, the responder is responsible for starting the message exchange sending GET\_VERSION to the responder, and it answers with all the supported SPDM versions.
Thereafter, they negotiate the common capabilities with the pair GET\_CAPABILITIES and CAPABILITIES, i.e., which type of SPDM requests are supported by each endpoint.
Then, to complete the connection context initialization, the requester sends their supported cryptographic algorithms with NEGOTIATE\_ALGORITHMS, then the responder choose one algorithm for each cryptographic operation (e.g., asymmetric, symmetric, and hash algorithms), and answers the request with ALGORTHMS response.
Hereafter, the exchanged messages depend on the capabilities negotiated beforehand.

Considering the communication with support for every request available, the requester sends GET\_DIGESTS to retrieve the certificates digests of the responder, willing to verify if it had been authenticated in previous communications.
If the received digests do not match with those stored in the requester cache, it is going to need the certificates in the responder memory, sending the GET\_CERTIFICATE request.
After selecting the desired certificates, the CHALLENGE request starts the authentication process in runtime.
Both endpoints build their own transcription of the exchanged messages, and the requester also sends a random value that the responder shall use together with its transcription to generate a signed response using its private key.
The requester receives the CHALLENGE\_AUTH response, checks the signature using the responder public key, then compares the received transcription with its own.
After a successful challenge, the responder can ask for the requester certificate, performing Basic Mutual Authentication.

This type of mutual authentication is deprecated, and has the sames steps of authenticating the responder: asks for digests and certificates, if necessary, and, then, challenges the target.
Nevertheless, the request messages coming from the responder as well as the response messages from the requester are encapsulated since a requester can only send requests and a responder, responses.

Subsequently, the requester sends the GET\_MEASUREMENTS request, which can be requested after the version, capabilities, and algorithms (VCA) phase.
Depending on the negotiated capabilities, measurements can be requested without authentication, however SPDM specification discourages this order to assure the responder identity because the GET\_MEASUREMENTS request clears the messages transcripts used for challenge.
Besides the order, when the requester receives an answer, it checks the message signature, and the measurements may be compared with the expected values if desired.

Finally, a secure session starts with the KEY\_EXCHANGE request, triggering the computation of a pair of ephemeral keys that is shared with KEY\_EXCHANGE\_RSP.
After receiving the new key, the endpoints are mutually authenticated before establishing a secure session, which is the advised way of performing mutual authentication since SPDM version 1.2.0.
This case of mutual authentication is named as Session-based Mutual Authentication by the SPDM specification~\cite{spdm-1.3.2:2024}.
Then, they complete the key negotiation with the message pair FINISH/FINISH\_RSP, enabling an encrypted communication by confirming the keys, and linking them to each endpoint identity.

\subsection{Trusted Platform Module} \label{subsec:tpm}

The security module, Trusted Platform Module (TPM), is mainly responsible for providing means to attest system devices.
It also offers the required cryptographic algorithms for device attestation and key generation.
Besides furnishing cryptographic functions, TPM can store keys, certificates, and device states~\cite{practical-tpm:2015}.

TPM has non-volatile indices to store two types of data: structured data defined by TPM architecture and non-structured data defined by the platform~\cite{practical-tpm:2015}.
In our proposal, secure boot uses the latter type of data to log SPDM authentication and measurement results for attestation purposes, i.e., they extend the index, and the memory region tied to the log cannot be erased.
The index extension consists of the following steps: retrieve the current hash value stored in the target index; compute the desired data hash; concatenate the retrieved value with the computed one; compute the hash of the concatenated data; and store this last value back in the target index~\cite{practical-tpm:2015}.

\begin{equation}\label{eq:hcrtm}
PCR_0[ALGO] = H_{ALGO}(0x4||H_{ALGO}(Data))
\end{equation}

Another resource available in TPM that our PoC uses is Core Root of Trust Measurement (CRTM), which computes the hash of the next booting stages and extend the Platform Configuration Register (PCR) responsible for platform initialization, i.e., PCR[0].
Although it is possible to use a software to compute the CRTM, our work uses TPM itself to compute the hash, which is named as Hardware CRTM (H-CRTM).
Equation \eqref{eq:hcrtm} demonstrates how TPM computes the H-CRTM using the desired algorithm before initializing all its features.
The concatenation of the hashed data with 0x4 indicates that the extension is performed by a hardware~\cite{practical-tpm:2015}.
%

\section{Proof-of-Concept} \label{sec:poc}

This section details the execution of our proof-of-concept, which, as explained in Section~\ref{sec:proposal}, is an emulated system that authenticates all PCIe and USB devices.
%

\subsection{Execution Flow} \label{subsec:execution-flow}

As soon as the emulator runs, it reads the OVMF code from the virtual flash memory to compute the H-CRTM before starting the system.
To verify the firmware integrity, the emulator compares the computed value extended at PCR[0] with the one stored in the variables section, which is computed beforehand. 
Nevertheless, since the pre-computed value can be tampered, we also store a signature in the variables section that is the H-CRTM signed with the Platform Key (PK) private key, which shall not be disclosed.
Therefore, our PoC guarantees that only an authorized firmware can run.

During Pre-EFI Initialization (PEI) stage, the UEFI firmware and TPM are mutually authenticated right after OVMF successfully locates the TPM.
Although our PoC assumes the TPM is reliable, similar to real systems behavior, and after verifying the firmware integrity we also trust the running firmware, we implement mutual authentication at this point for two reasons: to add redundancy and to implement a use case of SPDM in PEI.
The latter is useful for developers willing to authenticate hardware initialized during PEI (e.g., TPM, CPU, and memory) since TPM authentication with SPDM in our PoC provides an example design to base on.

Afterwards, during Driver Execution (DXE) stage, OVMF authenticates every PCIe and USB device willing to connect with the system during bus enumeration.
Similar to PEI stage, every device authenticates the firmware, thus implementing a zero trust architecture.
At DXE, the SPDM messages concerning certificates, challenge, and measurements extend the TPM's PCR as \cite{edk2-devicesecurity} for attestation purposes.

\begin{figure*}[!t]
    \centering
    \includegraphics[width=\textwidth,keepaspectratio]{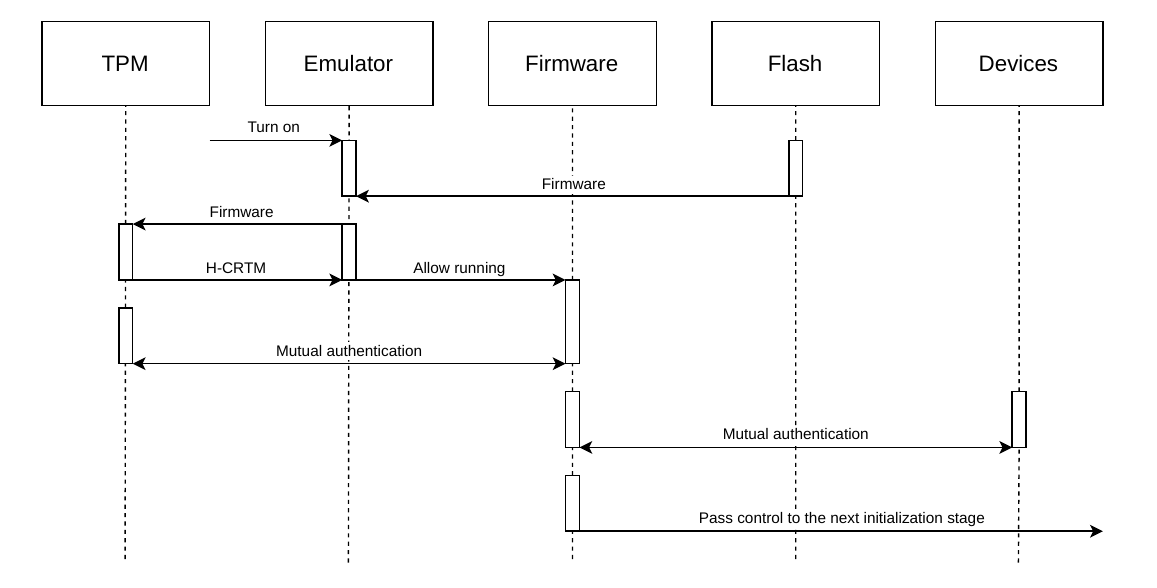}
    \caption{Communication between entities in our proposed system.}
    \label{fig:sec-boot-spdm}
\end{figure*}

Finally, OVMF finishes its initialization process with the boot selector and transfers control to the next stage, i.e., an OS.
Fig.~\ref{fig:sec-boot-spdm} shows the interactions between the system entities.

\subsection{Generating Certificates and H-CRTM} \label{subsec:certificates}

Platform Key (PK) and Key Exchange Key (KEK) are digital certificates to boot an OS with secure boot enabled, and we generate them using a custom tool written in C with GnuTLS library.
We choose this library due to its presence in QEMU, allowing the use of cryptographic functions without the need of adding another code tree to the emulator.
The tool performs the following steps:
\begin{enumerate}
    \item generates one RSA 2048 digital certificate, i.e., PK, which is the Certificate Authority (CA);
    \item generates one allowed database (DB) with the allowed signatures;
    \item generates one Certificate Sign Request (CSR) for an RSA 2048 certificate signed by PK, which results in KEK;
    \item computes H-CRTM with SHA-256; and
    \item signs H-CRTM with PK.
\end{enumerate}

In \cite{edk2-devicesecurity}, an EFI application stores the digital certificates in runtime variables, thus the user must run it at least once to have the certificates available, which requires shell interaction, however, when OVMF is built with secure boot, the shell is not enabled.
Because of this design, PEI authentication in \cite{edk2-devicesecurity} is only possible after running the emulation once to update the variables.
To overcome this restriction, we use a tool written in Python called ovmfvartool~\cite{ovmfvartool} that can analyze the EFI variables image and translate it to a YAML file as well as create an EFI variables image from a custom YAML file.
Therefore, we make a custom Python script to read the certificates and H-CRTM; fill the required EFI variable header with mandatory fields (e.g., GUID, variable name,  permissions, and optional fields such as date and time); and encode their data in base64.
%

\subsection{Emulator Verifications} \label{subsec:emulator-verifications}

OVMF has two main sections: variables and code.
The latter section is immutable, while the former is not as it stores certificates, allow and block lists, and other user and system defined variables.
Our PoC must not trust the code section, thus it verifies the firmware integrity before allowing its execution.
Therefore, we modify the TPM interface emulated by QEMU to read the flash memory and send the binary to the security module in order to compute the H-CRTM.

With the H-CRTM value computed during runtime, it is necessary to compare it with the expected one.
To make this comparison, we use the H-CRTM computed beforehand by a custom executable in the EFI variables as explained in Subsection~\ref{subsec:certificates}.
Nevertheless, since the variables section can be tampered, an attacker with enough knowledge about the verification process is able to modify the expected value.
Hence, to avoid this scenario, we sign the H-CRTM with the PK private key and store this signature in the EFI variables image.
In conclusion, if any verification fails, the emulation is terminated, and the firmware is not even executed.
%

\subsection{SPDM in QEMU} \label{subsec:spdm-in-qemu}

We insert the libspdm as a submodule inside QEMU code tree, enabling concurrently building and making available its Application Programming Interface (API) throughout the code.
Although using the libspdm as a submodule eases development, it prevents contributing directly to the emulator repository because this library uses tools, such as openSSL, with conflicting licenses.

With the API available, we write functions and a data structure, SpdmDev, to standardize the responder behavior, leveraging the code available in \cite{spdm-emu}, while adapting some parts to use our data structure.
As the responders have similar SPDM context configuration, we need to differentiate the SpdmDev instances, hence we define a linked list to store them.
This list allows searching for SpdmDev using a specific information tied to the device such as the Data Object Exchange (DOE) data structure that identifies PCIe devices, and USBDevice that identifies USB devices.
Regarding the SpdmDev for TPM, we insert an SpdmDev pointer in the TPM emulator own data structure, discarding any need to find it in the linked list, even though it is a member of the list.
%

\subsection{SPDM Communication for TPM} \label{subsec:tpm-spdm-comm}

There is an ongoing specification that defines the layout to encapsulate SPDM messages in TPM commands that our PoC uses~\cite{tpm-spdm:2025}.
It defines a header of two bytes (e.g., 0x8101 for clear messages and 0x8201 for secure messages), thus, since the message transport function is not yet implemented in libspdm, we write two functions to encode and decode SPDM messages for TPM.
These functions only manipulate the first six bytes of the TPM command header, i.e., tag and size, taking care to correctly assign the tag value based on the SPDM message type: clear or secure.
We also define two other transport functions that calls the encode and decode functions, which have the exact same logic of the existent transport functions in libspdm.
%

\subsection{PCIe Devices} \label{subsec:pcie-devices}

PCIe devices exchange messages using PCI DOE protocol, since the emulator already implements it, we only modify its data structure to store the address of the packet first header.
Every time the DOE mailbox receives data, it stores them using a variable that indicates the position of the next index in DOE write mailbox, which is a circular buffer requiring index manipulation to find the data packet.
Therefore, rapidly acquiring the address of the first header eases data manipulation in subsequent steps.

The protocol also requires a callback function to handle the received data packet.
We implement one that verifies if the first header carries an SPDM message, and, if it is, the function copies the content of the write mailbox to the buffer in the device SPDM context and call the libspdm method to dispatch the request processing.
Afterwards, the callback function asks libspdm for the response and copies it to the DOE read mailbox to make it available for the firmware.

Regarding the firmware, as discussed in Section~\ref{sec:related-work}, the code using SPDM is adapted from \cite{edk2-devicesecurity} and \cite{edk2-pci-doe}.
Besides updating outdated EDK II function calls, we do not modifying anything concerning PCIe authentication.
%

\subsection{USB Devices} \label{subsec:usb-devices}

To exchange SPDM messages between the hardware and firmware on the USB bus, we use the Management Component Transport Protocol (MCTP).
When OVMF receives the SPDM message, it does not know the packet data size beforehand, which would require a buffer with enough memory to hold the maximum SPDM configured message size, thus wasting system resources.
To solve this problem, we insert two bytes before the SPDM response, i.e., device-to-host direction, hence the driver shall ask for the first 2 bytes before receiving the remaining data.
On the other hand, when the transfer is host-to-device, the emulator already receives the exact total message size, for the firmware already has a method to send only the desired number of bytes to save system resources, thus there is no need to make any adaption.

Besides taking care of the system resources, there are multiple USB transfer types (e.g., control, isochronous, bulk, and interrupt), and the only one that every device must support is the control type, which configures the device before system usage.
Therefore, we modify only the control transfer handle function in QEMU to cover all USB devices.
Appendix~\ref{app:usb-msg-format} shows the USB header format our proposal implements.

Different from PCIe devices, the leveraged repositories do not have any implementation regarding USB, thus we copy and adapt their code logic.
Similarly, the device configuration is triggered when the firmware is enumerating the bus, and it detects a connection.
At this moment, OVMF installs the device security protocol with SPDM to authenticate and measure the peripheral, which only initializes its I/O functions of control type, before configuring it completely.
If the device successfully passes all the SPDM verifications, its initialization can proceed, otherwise the firmware skips this process and discards any memory region allocated for it.

The USB enumeration in OVMF happens after the System Management Mode (SMM) locks the memory to protect the system against unauthorized accesses, which triggers the hierarchy change in TPM, thus our DXE driver cannot extend the PCR after the lock event.
Then, we modify the order of the firmware execution, running the USB enumeration before the SMM lock.
This design does not damages the system security because, even though the USB device would technically be able to access memory regions that would be inaccessible, the SPDM authentication guarantees that only trusted devices can connect with the system.
Nevertheless, after the SMM lock event, no more USB devices can connect, hence the user must connect them before reaching this moment, otherwise the SPDM authentication will fail, and they will not be initialized.
%

\subsection{Mutual Authentication} \label{subsec:mut-auth}

Our PoC implements the SPDM standard until the KEY\_EXCHANGE request, however, in EDK II code tree, there are no functions to perform mutual authentication, basic or in secure session.
Therefore, in order to mutually authenticate the firmware and device, we copy the required functions from the devices secret library in libspdm to enable two features: sign the data with the requester private key and initialize a memory region with opaque data.
While the function providing the latter feature does not need any modification to work in EDK II, the function responsible for the former feature is altered to retrieve the private key from the EFI variables, which is different from libspdm that gets it from a file.
%

\section{Enabling SPDM in Other Components} \label{subsec:spdm-other-components}

It is possible to adapt other QEMU devices to stablish an SPDM communication.
For USB devices, no further modifications in new components are required as, in our PoC, every USB device is already capable of communicating using SPDM.
PCIe devices, on the other hand, can replicate the code of our PoC in other desired devices.
Regarding the modified OVMF, there is no need to change any code as the SPDM requester for both buses already supports any PCIe and USB device.

Devices with different buses can also use SPDM, yet the functions from our PoC do not contemplate them.
However, any developer willing to increase device support should follow the steps below:
\begin{enumerate}
    \item understand the beginning of the device connection logic;
    \item find a suitable moment to stablish an SPDM communication to authenticate the device, which should be after initializing its I/O capabilities and before completely initializing it;
    \item write encode/decode functions for message transport if not using MCTP or PCI DOE; and
    \item use the functions our PoC defines or create tailored ones to initialize SPDM and stablish a communication.
\end{enumerate}
%

 \section{Experiments} \label{sec:experiments}

We choose Linux as operating system for our PoC due to its code being open source and due to the existence of many educational materials and guides around it on the Internet.
Regarding the loader, we use the Linux EFI stub, which is a kernel as an EFI executable, simplifying the boot after OVMF as boot loaders such as GRUB are not required in UEFI~\cite{arch-efistub:2025}.
To use this method, it is only required to have a root file system with utilitarian binaries, and, to provide them, we use BusyBox because it produces a single compact executable with multiple tools~\cite{busybox:2024}.

This section discusses the PoC emulation in Subsection~\ref{subsec:customizing-emulation}; details the test scenarios in Subsection~\ref{subsec:test-scenarios}; describes the results of these scenarios in Subsection~\ref{subsec:running-scenarios}; presents the collected CPU metrics in Subsection~\ref{subsec:cpu-metrics}, and discusses them in Subsection~\ref{subsec:discussion}.
%

\subsection{Customizing Emulation} \label{subsec:customizing-emulation}

The emulation for our PoC discards the default devices QEMU instantiates to have more control on the system architecture, instantiating a PCIe hierarchy (e.g., solid state drive, network interface card, graphics processing unit, serial, and random number generator devices) and three USB devices (e.g., flash drive, mouse, and keyboard).
Alongside these components, we emulate TPM with the software \texttt{swtpm} and instantiate the TPM Interface Specification (TIS) backend in QEMU to communicate with the emulated hardware.
Besides these devices, we override the default BIOS with the custom OVMF loaded in two flash memories: one stores the code section and the other, the variables section.
Further details regarding the parameters to instantiate our PoC is in Appendix~\ref{app:emu-details}.
%

\subsection{Test Scenarios} \label{subsec:test-scenarios}

We describe four different test scenarios: one ideal and three with expected errors.
In the ideal one, both firmware verifications occurs successfully, and the devices have permission to connect with the system.
The second scenario fails during firmware integrity verification; the third one fails when verifying firmware authenticity; and the last one fails while authenticating devices.

Table~\ref{tab:test-scenarios} is an overview of these scenarios.
We use ``\faTimes'' when we expect a failure and ``\faCheck'' when we expect a successful operation.
In Scenarios 2 and 3, the permission to connect a device is irrelevant, for the emulation terminates before authenticating the devices.
Similarly, Scenario 2 does not reach firmware authentication, thus this field is also irrelevant.

\begin{table*}[!t]
    \centering
    \caption{Test scenarios overview.}
    \begin{tabular}{|c|c|c|c|}
        \hline
         \multirow{3}{8em}{\centering \textbf{Test Scenario}} & \multirow{3}{8em}{\centering \textbf{Firmware Integrity}} & \multirow{3}{8em}{\centering \textbf{Firmware Authenticity}} & \multirow{3}{8em}{\centering \textbf{Devices Allowed}} \\
         &&&\\
         &&&\\
         \hline
         \textbf{1} & \faCheck & \faCheck & \faCheck \\
         \hline
         \textbf{2} & \faTimes & \faMinus & \faMinus \\
         \hline
         \textbf{3} & \faCheck & \faTimes & \faMinus \\
         \hline
         \textbf{4} & \faCheck & \faCheck & \faTimes \\
         \hline
    \end{tabular}
    \label{tab:test-scenarios}
\end{table*}

\subsection{Scenario Execution} \label{subsec:running-scenarios}

To run the Scenario 2, we rebuild OVMF and do not update the expected H-CRTM in the EFI variables, which returns the message ``\texttt{SECURITY ERROR - FIRMWARE MODIFIED}'' when QEMU compares the hashes.
When running Scenario 3, on the other hand, we update the variables with the current OVMF H-CRTM, yet we sign it with an unrecognized private key, thus the firmware successfully passes the integrity check and fails during authentication, returning the message ``\texttt{Error verifying firmware hash - Public Key signature verification has failed}''.
As expected, these two Scenarios terminate the emulation right after printing their respective messages.

Regarding Scenario 4, it must fail when trying to authenticate disallowed devices, thus the firmware must report an authentication fail during CHALLENGE, i.e., error code 0x80000030.
To force the error in this scenario, we tamper the expected certificate that is in ``RequesterSpdmCertChain'' variable, preventing the connection since the target devices do not send the authorized certificate to the firmware.
In this Scenario, all the PCIe and USB devices become unusable as the emulation instances them with the same certificate as discussed in Section~\ref{sec:poc}, hence the OS is not booted because the NVMe mass storage device is not initialized, and every authentication attempt returns the message ``\texttt{Authentication - 0x80000030}''.

Scenario 1 runs as expected without resulting in any error, and it is used in subsequent sections to analyze the CPU running the emulation.
%

\subsection{CPU Metrics} \label{subsec:cpu-metrics}

We count the number of instructions and CPU cycles until BusyBox Init finishes its initialization process and is ready for user login.
The only part of the boot that has SPDM is the OVMF, thus the OS does not know anything about the authentication process, i.e., any unauthorized device keeps uninitialized when Linux boots, which can initialize the device afterwards, however this is outside the scope of our work.
We do not measure Scenarios 2 and 3 metrics because they do not run any CPU instructions, and, similarly, Scenario 4 cannot even boot Linux since the NVMe is not initialized as discussed in Subsection~\ref{subsec:running-scenarios}.

To gather the metrics for our emulated system, we use a host machine with an AMD Ryzen 7 2700 and Linux with OS virtualization enabled, i.e., Kernel Virtual Machine (KVM).
If both guest and host machine architectures match, QEMU exposes the real CPU for the guest machine, and it can run the instructions on the host CPU with KVM help, hence, to use this feature, we must configure the emulator with \texttt{--enable-kvm} parameter and also execute it with \texttt{-enable-kvm} option on the command line.
At the moment the OS asks for user credentials, we force the machine shutdown without requiring any user interaction, and we collect the number of instructions and CPU cycles using ``perf'', which has a KVM option to count the desired metrics for the guest machine.
With this technique, we run the emulated system 10 thousand times for the Scenario 1 and for the system without SPDM each to compare them.

\begin{figure}[!t]
    \centering
    \includegraphics[width=\linewidth]{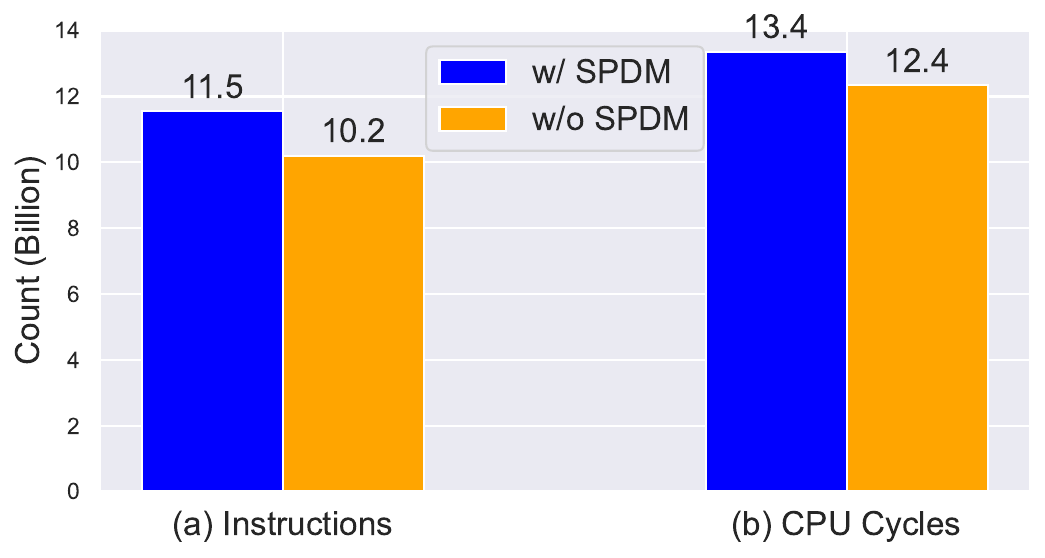}
    \caption{Comparison of the average number of (a) instructions and (b) CPU cycles when running the system with and without SPDM.}
    \label{fig:avg-cmp}
\end{figure}

After collecting the number of instructions and CPU cycles, we note the data has measures with over 100 billion instructions and even 1 trillion CPU cycles, which are not consistent across the majority of the data.
As a result, we compute the z-score for both metrics with and without SPDM to remove the ones we classify as outliers, i.e., those with z-score greater than 2 and less than -2, which represent approximately 2.2\% of the total.
Afterwards, we compute the average for both measures resulting in approximately 11.5 billion instructions when using SPDM and 10.2 billion instructions when not using it, deviating 0.3 billion instructions and 0.4 billion instructions, respectively.
CPU cycles, on the other hand, resulted in 13.4 billion cycles when using SPDM and 12.4 billion when not using it, showing greater deviations than the instruction count with 3 billion cycles and 4.1 billion cycles, respectively.
Fig.~\ref{fig:avg-cmp} shows the comparison of the averages between the two scenarios, and Appendix~\ref{app:metrics} has tables summarizing the collected data.

\subsection{Discussion} \label{subsec:discussion}

We can compute the total CPU time by dividing the average of CPU cycles by the CPU frequency.
To get the frequency, we use the ``lscpu'' command, which returns 1.55 GHz as the minimum and 3.20 GHz as the maximum.
Afterwards, we divide the average of CPU cycles with and without SPDM resulting in the minimum CPU time of 4.2 s and 3.9 s, respectively, and the maximum of 8.6 s and 8 s, respectively. 
The emulated system with SPDM authentication when compared to the one without SPDM authentication has an increase in CPU time, cycles and instructions of approximately 8\%, 13\%, and 8\%, respectively.
This increase in the CPU metrics can be acceptable since it only happens during the UEFI firmware execution, while providing protection against unauthorized connections of PCIe and USB devices. 

Concerning the possible attacks described in Section~\ref{sec:attacker-model}, our proposal can successfully protect the system against them as they depend on malicious device connection.
However, as discussed is Section~\ref{sec:attacker-model}, our PoC effectiveness depends on the protection of the private key used to sign the certificates stored in the authorized devices, otherwise the attacker is able to counterfeit an authorized device.
Therefore, with the private key safely stored, even if an attacker with the highest PA level replaces a component, our PoC would not allow its connection because its certificate would not be signed by the expected private key.
Similarly, unaware users could connect an unauthorized device to the system physical port, yet the UEFI firmware would not initialize it, which effectively defends the system against PA-3 attacks.
%

\section{Final Remarks} \label{sec:final-remarks}

Our work proposes a UEFI system able to authenticate PCIe and USB devices with SPDM during boot, and we develop a proof-of-concept with open source tools, making it available in a public repository\footnote{\url{https://github.com/htafr/uefi-spdm/tree/0d8a895}}.
The provided resources are able to increase development speed since institutions demanding high security can leverage the PoC to tailor their own UEFI system with SPDM according to their requirements.
%
%
Despite the contribution, our PoC evaluates the overhead of CPU instructions and cycles, noting an increase of 8\% in CPU time and cycles, and of 13\% in CPU instructions.

There are three future work directions: accurate hardware measurements, performance enhancements, and improving security after OS initialization.
Regarding the first, since there is no commercially available SPDM enabled hardware, we plan to develop a custom one using Field Programmable Gate Array (FPGA) to evaluate our PoC.
Concerning performance, there is a potential research topic about persisting the SPDM context, instead of destroying it, to use it on subsequent boot phases, which would save resources as the whole SPDM communication would not be repeated when the OS, capable of communicating with SPDM, is initializing.
In conclusion, for further security, the SPDM context could be persisted in the TPM instead of the main memory, saving the system resources and also protecting it from tampering attempts.

\section*{Acknowledgment}

This work was supported by FAPESP (2020/09850-0), CNPq (307732/2023-1), and CAPES (Finance Code 001).

\printbibliography

\begin{appendices}

\section{USB Control Transfer Message Format} \label{app:usb-msg-format}

The bitmap ``bmRequestType'' depends solely on the data transfer direction, assuming 0x0 and 0x80 when it is host-to-device and device-to-host, respectively.
We arbitrarily choose 0x32 as ``bRequest'' to represent SPDM requests, taking due care to not choose another value already in use to identify other requests.
``wValue'' is always 0x0 since the responder is a USB device and not an interface or an endpoint.
Lastly, ``wIndex'' carries the data packet offset, and ``wLength'', the data packet size.

Tables~\ref{tab:usb-req-format-dev2host} and \ref{tab:usb-req-format-host2dev} show the data packet headers when the direction is device-to-host and host-to-device, respectively.

\begin{table}[!b]
  \caption{Device-to-host control transfer header.}
  \begin{center}
    \begin{tabular}{*{4}{|c}|}
      \hline
      \textbf{Offset}& \textbf{Field}& \textbf{Size}& \textbf{Value} \\
      \hline
      0& wTotalSize& 2& SPDM message size \\
      \hline
      2& bmRequestType& 1& 0x80 \\
      \hline
      3& bRequest& 1& 0x32  \\ 
      \hline
      5& wValue& 2& 0x0 \\
      \hline
      7& wIndex& 2& Bytes offset \\
      \hline
      9& wLength& 2& Number of bytes \\
      \hline
    \end{tabular}
  \label{tab:usb-req-format-dev2host}
  \end{center}
\end{table}

\begin{table}[!b]
  \caption{Host-to-device control transfer header.}
  \begin{center}
    \begin{tabular}{*{4}{|c}|}
      \hline
      \textbf{Offset}& \textbf{Field}& \textbf{Size}& \textbf{Value} \\
      \hline
      0& bmRequestType& 1& 0x0 \\
      \hline
      1& bRequest& 1& 0x32  \\ 
      \hline
      2& wValue& 2& 0x0 \\
      \hline
      4& wIndex& 2& Bytes offset \\
      \hline
      6& wLength& 2& Number of bytes \\
      \hline
    \end{tabular}
  \label{tab:usb-req-format-host2dev}
  \end{center}
\end{table}

\section{Emulation Details} \label{app:emu-details}

To create the emulator instance for our PoC, we use the following QEMU parameters~\cite{qemu-docs:2024}:
\begin{itemize}
    \item \texttt{bios none}: indicate to not load the default firmware, i.e., SeaBIOS;
    \item \texttt{nodefaults}: discard QEMU default devices (e.g., serial port, parallel port, virtual console, monitor device, etc.;
    \item \texttt{device}: instance a new device;
    \begin{itemize}
        \item \texttt{bus}: specify the bus ID to connect the device;
    \end{itemize}
    \item \texttt{drive}: define a new drive to tie to the block device backend;
    \item \texttt{usb}: enable USB emulation;
    \item \texttt{chardev}: instance a character device;
    \begin{itemize}
        \item \texttt{socket}: instance a character device of socket type;
        \item \texttt{id}: specify any string to identify the character device; and
        \item \texttt{path}: specify the path to create the socket file.
    \end{itemize}
    \item \texttt{tpmdev}: instance a TPM backend device;
    \begin{itemize}
        \item \texttt{emulator}: indicate the TPM backend device is an emulator;
        \item \texttt{id}: specify any string to identify the character device; and
        \item \texttt{chardev}: specify the character device ID to connect the device backend.
    \end{itemize}
\end{itemize}

Using these parameters, we instance the following PCIe devices:
\begin{itemize}
    \item \texttt{device pcie-root-port}: create one PCIe root port attached to the root bus, creating a PCIe only hierarchy;
    \begin{itemize}
        \item \texttt{id}: identify the bus with any string; and
        \item \texttt{chassis}: identify the \texttt{chassis} with any string;, which must be a unique \texttt{chassis}, \texttt{slot} pair, and the latter is 0 by default.
    \end{itemize}
    \item \texttt{device nvme}: instance an NVMe mass storage device;
    \begin{itemize}
        \item \texttt{serial}: define a custom serial number; and
        \item \texttt{drive}: specify the \texttt{drive} ID tied to this device.
    \end{itemize}
    \item \texttt{device virtio-serial-pci}: instance a PCI virtual I/O serial device;
    \item \texttt{device virtio-rng-pci}: instance a PCI virtual random number generator (RNG) device;
    \item \texttt{device virtio-gpu-pci}: instance a PCI virtual graphic processing unit (GPU);
    \item \texttt{device e1000e}: instance an e1000 network interface card (NIC);
    \begin{itemize}
        \item \texttt{netdev}: specify the network device it is tied to.
    \end{itemize}
\end{itemize}

The following are the USB devices our emulation use:
\begin{itemize}
    \item \texttt{device qemu-xhci}: instance an XHCI controller on machines with PCI, which supports any USB version;
    \begin{itemize}
        \item \texttt{id}: identify the bus with any string.
    \end{itemize}
    \item \texttt{device usb-storage}: instance a USB storage device;
    \begin{itemize}
        \item \texttt{drive}: specify the \texttt{drive} ID tied to this device; and
        \item \texttt{removable}: define if the device is removable, i.e., USB stick.
    \end{itemize}
    \item \texttt{device usb-kbd}: instance a USB keyboard device.
    \item \texttt{device usb-mouse}: instance a USB mouse device;
\end{itemize}

We emulate TPM using swtpm~\cite{swtpm-docs}, which can emulate both versions: 1.2 and 2.0.
This software is essential to use the security module interface that the emulation provides, and, for it to work together with QEMU, we must instantiate two devices: \texttt{chardev socket} and \texttt{tpmdev}~\cite{qemu-docs:2024}.
The former routes the data to a file that has read and write permissions for QEMU and swtpm, and the latter is the TPM backend to communicate with the emulated security module.
To get QEMU and swtpm working together, we use the following emulator parameters:
\begin{itemize}
    \item \texttt{device tpm-tis}: instance a TPM interface device of type TPM Interface Specification (TIS);
    \begin{itemize}
        \item \texttt{tpmdev}: specify the TPM backend device ID.
    \end{itemize}
\end{itemize}

Alongside TIS, we run the swtpm and use the following parameters to customize the emulated TPM:
\begin{itemize}
    \item \texttt{tpm2}: instance a TPM 2.0;
    \item \texttt{d}: daemonize the instance;
    \item \texttt{tpmstate}: specify the path where swtpm is going to store the TPM state;
    \item \texttt{ctrl}: specify the controller type and socket file path, which is \texttt{unixio} as Linux manages the socket file that QEMU creates with \texttt{chardev socket}; and
    \item \texttt{log}: set the log level as 20, which is the commonly used value for verbose output~\cite{swtpm-docs,qemu-docs:2024}.
\end{itemize}

The remaining required devices are the flash memories to store the OVMF, which we instance using the \texttt{drive} QEMU parameter.
We use the following options to customize it:
\begin{itemize}
    \item \texttt{if}: specify the interface, and, for flash memory, we set it as \texttt{pflash};
    \item \texttt{unit}: specify a numerical value to identify the flash memory;
    \item \texttt{format}: specify the image format, which is \texttt{raw} for the OVMF binaries;
    \item \texttt{file}: specify the image path; and
    \item \texttt{readonly}: mark the flash memory as read only if the value is \texttt{on}.
\end{itemize}

The \texttt{readonly} parameter prevents QEMU from writing the binary, which is the desired behavior for the code section of OVMF.
On the other hand, the EFI variables image must have write permissions, however we use the \texttt{global} parameter with options that does not persist the modifications in the binary after the emulation terminates.
We use the following options with this parameter:
\begin{itemize}
    \item \texttt{driver}: select the drive to apply a property (e.g., \texttt{cfi.pflash01} selects the flash memory of unit 1);
    \item \texttt{property}: select the property (e.g., \texttt{secure} prevents QEMU from modifying the original image); and
    \item \texttt{value}: set the property value (e.g., \texttt{on} or \texttt{off}).
\end{itemize}

\section{Metrics} \label{app:metrics}

Tables~\ref{tab:nospdm-metrics} and \ref{tab:spdm-metrics} summarizes the data collected from the experiments detailed in Section~\ref{sec:experiments}.

\begin{table}[!t]
    \centering
    \caption{Booting system \textbf{without} SPDM.}
    \label{tab:nospdm-metrics}
    \begin{tabular}[\textwidth]{r r}
        \hline
        \textbf{Instructions (billion)} & $10.20 \pm 0.43$ \\
        \textbf{CPU Cycles (billion)} & $12.35 \pm 4.09$ \\
        \textbf{IPC} & $0.83 \pm 0.06$ \\
        \hline
    \end{tabular}
\end{table}

\begin{table}[!t]
    \centering
    \caption{Booting system \textbf{with} SPDM.}
    \label{tab:spdm-metrics}
    \begin{tabular}[\textwidth]{r r}
        \hline
        \textbf{Instructions (billion)} & $11.55 \pm 0.33$ \\
        \textbf{CPU Cycles (billion)} & $13.36 \pm 3.00$ \\
        \textbf{IPC} & $0.87 \pm 0.06$ \\
        \hline
    \end{tabular}
\end{table}

\vfill
\end{appendices}

\end{document}